\batchmode
\makeatletter
\def\input@path{{"C:/Trabajo laptop/Mis articulos/Finished/Fusion 2022/Analysis adaptive labelled birth/"}}
\makeatother
\documentclass[english]{IEEEtran}
\usepackage[T1]{fontenc}
\usepackage[latin9]{inputenc}
\usepackage{amsmath}
\usepackage{amsthm}
\usepackage{amssymb}
\usepackage{graphicx}
\PassOptionsToPackage{normalem}{ulem}
\usepackage{ulem}

\makeatletter

\providecommand{\tabularnewline}{\\}

\theoremstyle{plain}
\newtheorem{thm}{\protect\theoremname}
\theoremstyle{plain}
\newtheorem{lem}[thm]{\protect\lemmaname}

\usepackage{cite} 
\usepackage[margin=8pt,font=footnotesize]{caption}
\usepackage{algorithm}
\usepackage{algpseudocode}
\usepackage{amsmath}  % You need this for the math
\allowdisplaybreaks

\@ifundefined{showcaptionsetup}{}{%
 \PassOptionsToPackage{caption=false}{subfig}}
\usepackage{subfig}
\makeatother

\usepackage{babel}
\providecommand{\lemmaname}{Lemma}
\providecommand{\theoremname}{Theorem}

\begin{document}
\title{A comparison between PMBM Bayesian track initiation and labelled RFS
adaptive birth}
\author{Ángel F. García-Fernández$^{\star}$, Yuxuan Xia$^{\circ}$, Lennart
Svensson$^{\circ}$\\
{\normalsize{}$^{\star}$Dept. of Electrical Engineering and Electronics,
University of Liverpool, United Kingdom}\\
{\normalsize{}$^{\star}$ARIES Research Centre, Universidad Antonio
de Nebrija, Spain}\\
{\normalsize{}$^{\circ}$Dept. of Electrical Engineering, Chalmers
University of Technology, Sweden}\\
{\normalsize{}Emails: angel.garcia-fernandez@liverpool.ac.uk, firstname.lastname@chalmers.se}}

\maketitle
\thispagestyle{empty}
\begin{abstract}
This paper provides a comparative analysis between the adaptive birth
model used in the labelled random finite set literature and the track
initiation in the Poisson multi-Bernoulli mixture (PMBM) filter, with
point-target models. The PMBM track initiation is obtained via Bayes'
rule applied on the predicted PMBM density, and creates one Bernoulli
component for each received measurement, representing that this measurement
may be clutter or a detection from a new target. Adaptive birth mimics
this procedure by creating a Bernoulli component for each measurement
using a different rule to determine the probability of existence and
a user-defined single-target density. This paper first provides an
analysis of the differences that arise in track initiation based on
isolated measurements. Then, it shows that adaptive birth underestimates
the number of objects present in the surveillance area under common
modelling assumptions. Finally, we provide numerical simulations to
further illustrate the differences. 
\end{abstract}


\section{Introduction}

Multiple target tracking (MTT) is the process of estimating target
trajectories based on noisy sensor data, such as radar and sonar,
and has a wide array of applications, including defence, automotive
systems and maritime traffic monitoring \cite{Blackman_book99}. Three
popular frameworks to solve the multi-target tracking problem are
multiple hypothesis tracking \cite{Chong19}, joint probabilistic
data association \cite{Fortmann83} and random finite sets (RFS) \cite{Mahler_book14}.
These apparently different approaches are closely related to each
other, with links established in \cite{Williams15b,Brekke18}. 

In the RFS framework, we require probabilistic models for target birth,
dynamics and death, as well as sensor measurements. For the standard
point target dynamic and measurement models, and a Poisson point process
(PPP) birth model, the posterior density of the set of targets, and
also the set of trajectories, is a Poisson multi-Bernoulli mixture
(PMBM) \cite{Williams15b,Angel18_b,Granstrom18}. If the birth model
is multi-Bernoulli, the posterior density of the set of targets and
the set of trajectories is a multi-Bernoulli mixture (MBM) \cite{Angel19_e,Xia19_b}.
Both PMBM and MBM filters have hidden/latent variables \cite{Dempster77,Pitt99},
e.g., representing the underlying data associations and the Bernoulli
birth component for multi-Bernoulli birth. For multi-Bernoulli birth,
one can make the latent variable that represents time of birth and
Bernoulli component explicit in the posterior to form a labelled RFS.
The resulting filter is a labelled MBM filter, whose recursion is
analogous to the MBM filter \cite{Angel18_b}. The MBM filter (labelled
or not) can be written with hypotheses with deterministic target existence
(MBM$_{01}$ filter), with an exponential increase in the number of
global hypotheses \cite{Angel18_b}. When labelled, the MBM$_{01}$
filter corresponds to the widely-used $\delta$-generalised labelled
multi-Bernoulli ($\delta$-GLMB) filter \cite{Vo14}. A computationally
faster approximation of the $\delta$-GLMB filter is the labelled
multi-Bernoulli (LMB) filter \cite{Reuter14}.  

\begin{figure}
\begin{centering}
\subfloat[]%
{\begin{centering}
\includegraphics[scale=0.9]{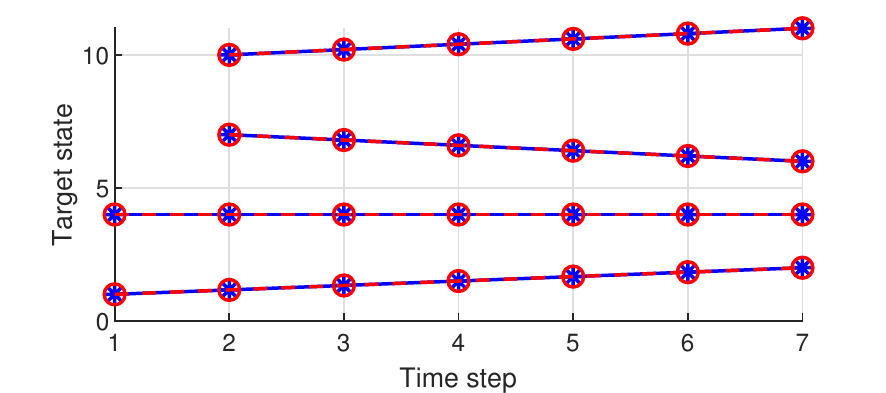}
\par\end{centering}

}
\par\end{centering}
\begin{centering}
\subfloat[]%
{\begin{centering}
\includegraphics[scale=0.9]{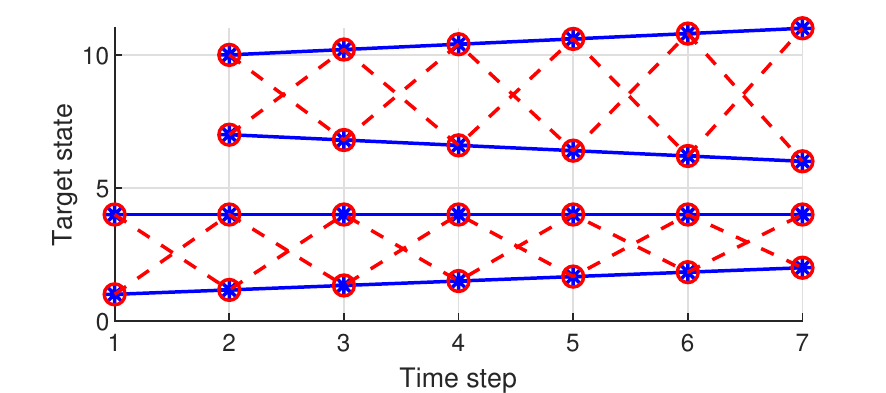}
\par\end{centering}

}
\par\end{centering}
\begin{centering}
\subfloat[]%
{\begin{centering}
\includegraphics[scale=0.9]{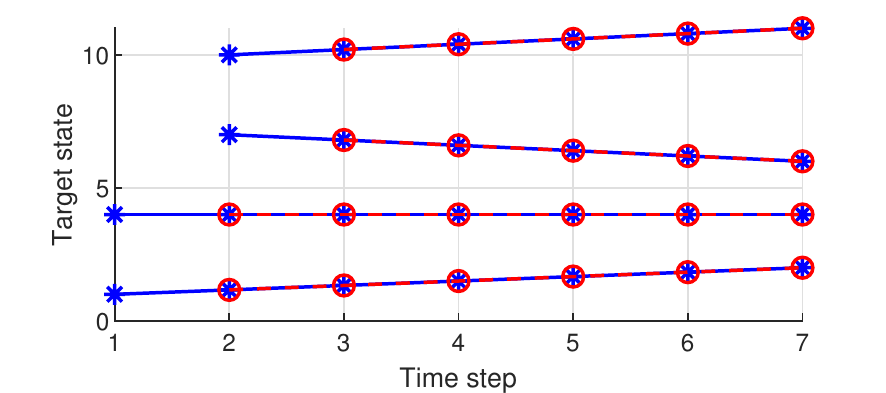}
\par\end{centering}

}
\par\end{centering}
\caption{\label{fig:Track_switching}Estimated set of trajectories with $p^{D}=1$,
$p^{S}=1$, accurate measurements, $\overline{\lambda}^{C}\rightarrow0$
and an IID cluster birth process that covers the whole surveillance
area, allowing more than one target to be born at each time. Blue
lines: ground truth set of trajectories. Red dashed line: estimated
set of trajectories. Each target cannot move more than 0.5 units at
each time step. Subfigure (a): Bayesian solution. Subfigure (b): Track
switching for targets born at the same time step. Subfigure (c): Misdetection
of targets at the birth time step. Trajectory PMBM and trajectory
Poisson multi-Bernoulli (PMB) filters provide (a). The PMBM and PMB
filters with sequential track building based on auxiliary variables
\cite{Angel20_e} also provide (a). The filtering densities of the
labelled MBM, $\delta$-GLMB and LMB filters do not have information
to distinguish between situations (a) and (b), though one can apply
the dynamic model to the estimated target states to choose (a). Labelled
filters with adaptive birth provide (c) if we select the parameters
as $r_{B,max}=1$, $\overline{\lambda}_{2}^{B}\protect\geq2$ and
$\overline{\lambda}_{3}^{B}\protect\geq2$.}

\end{figure}

The $\delta$-GLMB and LMB filters work well to estimate target states
and trajectories when the multi-Bernoulli birth process allows at
maximum one target to be born in a specific location. However, they
have difficulties to deal with an independent and identically distributed
(IID) cluster birth process covering a large area, in which more than
one target may be born at the same time step, for instance, a PPP.
In this case, to run the $\delta$-GLMB and LMB filters, we can approximate
the PPP as multi-Bernoulli with a sufficient number of components
to cover the cardinality distribution, and setting the spatial distribution
of the Bernoulli components as in the PPP. The challenge is that these
filters would require propagating a large number of global hypotheses
to convey the relevant information \cite[App. D]{Angel20_e}. In addition,
in this case, the estimated trajectories by the $\delta$-GLMB and
LMB filters are not always satisfactory for targets born at the same
time step, see Figure \ref{fig:Track_switching} and \cite[Ex. 2]{Angel20_b}.
This problem could be partially solved by partitioning the surveillance
area into a grid, and allowing at maximum one birth with a given label
in each cell \cite{Angel13}. 

Adaptive birth (also called measurement-driven birth) is the most
widely-used approach in the labelled RFS literature to deal with large
uncertainty in the birth process \cite{Reuter14,Nguyen21,Nguyen21b,Olofsson17,Moratuwage18,Li18,Li19c,Ong22,Reuter13,Do22}.
Here, each measurement creates a (labelled) Bernoulli birth component
at the following time step. This approach generally works well and
mimics track initiation in the PMBM filter, in which each measurement
creates a new track, represented as a Bernoulli component. However,
in the PMBM filter, the probability of existence and single-target
density of a new Bernoulli are obtained via Bayes' rule. In adaptive
birth, the single-target density of this Bernoulli component is user-defined
and the probability of existence depends on the probability that the
measurement is associated to previous targets, also adding a user-defined
threshold. While adaptive birth can work well for target and trajectory
estimation, the $\delta$-GLMB filter does no longer provide a recursion
to calculate the posterior in closed-form. In particular, target births
are dependent on past measurements, which does not agree with assumptions
required in the $\delta$-GLMB filter derivation \cite{Vo13}.

Adaptive birth has also been adjusted for extended targets \cite{Beard16},
merged measurements \cite[Fig. 3]{Beard15}, simultaneous localisation
and mapping \cite{Deusch15}, and combined with the standard multi-Bernoulli
birth \cite{Kim19}. There are also other variations \cite{Beard20}\cite{Trezza22}.
Another approach to the deal with this shortcoming of the $\delta$-GLMB
filter is provided in \cite{Legrand18}. It should also be mentioned
that, in \cite{Ristic12}, what is referred to adaptive birth is used
in a sequential Monte Carlo probability hypothesis density (PHD) filter
implementation. In \cite{Ristic12}, the birth intensity is actually
uniform, and a measurement-driven importance sampling is used for
drawing particles \cite{Gustafsson10}. 

In this paper, we compare the standard adaptive birth method for labelled
RFSs with the track initiation in the PMBM filter. To do so, we analyse
the differences in track initiation from isolated measurements using
both approaches. We also provide an analysis on the expected number
of targets for both methods. Finally, we compare both approaches via
simulations.

\section{MTT models and track initiation}

We first provide some background on the standard models for RFS filtering
and the filtering recursions in Section \ref{subsec:Bayesian-filtering-recursions}.
We then review Bernoulli track initiation in PMBM filters in Section
\ref{subsec:PMBM-Bernoulli-track-initialisation}. We then review
the adaptive birth model for labelled RFSs in Section \ref{subsec:Adaptive-birth-model}.

\subsection{Bayesian filtering recursions\label{subsec:Bayesian-filtering-recursions}}

In multi-target filtering, we are interested in computing the posterior
density of the set $X_{k}$ of targets at time step $k$, which is
the density of the set of targets given all past and current measurements.
At each time step, each target $x\in X_{k}$ is detected with probability
$p^{D}\left(x\right)$ and generates a measurement with conditional
density $l\left(\cdot|x\right)$. The set $Z_{k}$ of measurements
is the union of the set of target-generated measurements and the set
of clutter measurements, which are distributed according to a PPP
with intensity $\lambda^{C}\left(z\right)$. 

Each target $x\in X_{k}$ survives to the next time step with probability
$p^{S}\left(x\right)$ with a transition density $g\left(\cdot|x\right)$.
The set $X_{k+1}$ of targets is the union of the set of surviving
targets and the set of new born targets at time step $k+1$, which
is independent of the set of surviving targets. There are two standard
models for target birth: the PPP birth model, with intensity $\lambda_{k}^{B}\left(\cdot\right)$,
and the multi-Bernoulli birth model. With a PPP birth model, the posterior
is a PMBM, and with multi-Bernoulli birth model, the posterior is
a MBM, which can also be written in MBM$_{01}$ form \cite{Angel18_b,Angel19_e}. 

With multi-Bernoulli birth model, one can write the hidden variable
$\ell=(k,l)$, where $k$ is the time of birth and $l$ is an index
that represents a Bernoulli, explicit in the target state, such that
$x=\left(x',\ell\right)$, where $x'$ is the dynamic state without
label and $\ell$ is the target label \cite[Sec. IV]{Angel18_b},
and define $g\left(y',\ell_{y}|x',\ell_{x}\right)=g\left(y'|x'\right)\delta_{\ell_{x}}\left[\ell_{y}\right]$.
In this case, the recursion to compute the posterior is analogous
to the MBM case, but with labelled target states. The MBM$_{01}$
filter with labels corresponds to the $\delta$-GLMB filter.

\subsection{PMBM Bernoulli track initiation\label{subsec:PMBM-Bernoulli-track-initialisation}}

We proceed to explain how Bernoulli tracks are initialised in PMBM
filtering. Given $Z_{k}=\left\{ z_{k}^{1},...,z_{k}^{m_{k}}\right\} $,
each measurement generates a new Bernoulli obtained via Bayes' rule.
The Bernoulli created by measurement $z_{k}^{p}$ has probability
of existence and single-target density \cite{Williams15b,Angel18_b}
\begin{align}
r_{k|k} & =\frac{\left\langle \lambda_{k|k-1},p^{D}l\left(z_{k}^{p}|\cdot\right)\right\rangle }{\left\langle \lambda_{k|k-1},p^{D}l\left(z_{k}^{p}|\cdot\right)\right\rangle +\lambda^{C}\left(z_{k}^{p}\right)}\label{eq:r_k_PMBM}\\
p_{k|k}\left(x\right) & =\frac{p^{D}\left(x\right)l\left(z_{k}^{p}|x\right)\lambda_{k|k-1}\left(x\right)}{\left\langle \lambda_{k|k-1},p^{D}l\left(z_{k}^{p}|\cdot\right)\right\rangle }\label{eq:p_k_PMBM}
\end{align}
where $\lambda_{k|k-1}\left(\cdot\right)$ is the intensity of the
PPP representing undetected targets in the PMBM predicted density,
and $\left\langle f,g\right\rangle =\int f\left(x\right)g\left(x\right)dx$.

We can then see that the Bayesian approach sets the probability of
existence weighting two hypotheses: the probability that the measurement
has been generated by a potentially undetected target or clutter.
The single-target density takes into account a possibly state-dependent
probability of detection, and the state information on undetected
targets. 

\subsection{Adaptive birth model for labelled RFSs\label{subsec:Adaptive-birth-model}}

We proceed to explain how Bernoulli births are set in adaptive birth
\cite{Reuter14}. Adaptive birth does not use the (labelled) multi-Bernoulli
birth mentioned in Section \ref{subsec:Bayesian-filtering-recursions}.
First, it requires knowledge of the expected number of targets at
time step $k$, which we denote by $\overline{\lambda}_{k}^{B}$.
Note that $\overline{\lambda}_{k}^{B}$ can be obtained from the standard
birth models. For PPP birth, we have \cite{Mahler_book14}
\begin{align}
\overline{\lambda}_{k}^{B} & =\int\lambda_{k}^{B}\left(x\right)dx.
\end{align}
For multi-Bernoulli birth with $n_{k}^{b}$ potential targets, each
with a probability of existence $r_{k}^{l}$, we have
\begin{align}
\overline{\lambda}_{k}^{B} & =\sum_{l=1}^{n_{k}^{b}}r_{k}^{l}.
\end{align}

Given $Z_{k}=\left\{ z_{k}^{1},...,z_{k}^{m_{k}}\right\} $, the adaptive
birth model creates a (labelled) multi-Bernoulli with $m_{k}$ Bernoulli
components at time step $k+1$. The Bernoulli created by $z_{k}^{p}\in Z_{k}$
has a user-defined single-target density $p_{B,k+1}^{(\ell)}\left(x|z_{k}^{p}\right)$
and probability of existence
\begin{align}
\hat{r}_{B,k+1}^{(\ell)}\left(z_{k}^{p}\right) & =\min\left(r_{B,max},\frac{1-r_{U,k}\left(z_{k}^{p}\right)}{\sum_{j=1}^{m_{k}}\left(1-r_{U,k}\left(z_{k}^{j}\right)\right)}\overline{\lambda}_{k+1}^{B}\right)\label{eq:r_b}
\end{align}
where $r_{B,max}\in[0,1]$ is a user-defined parameter that sets a
maximum to the existence probability of the new Bernoulli components,
and $r_{U,k}\left(z\right)$ is the probability that a measurement
$z$ is associated to a target in the previous $\delta$-GLMB/LMB
hypotheses. In mathematical terms, we have
\begin{align}
r_{U,k}\left(z\right) & =\sum_{\left(I_{+},\theta\right)}1_{\theta}\left(z\right)w_{k}^{\left(I_{+},\theta\right)}\label{eq:r_u}
\end{align}
where $1_{\theta}\left(z\right)$ is one if $z\in\theta$ and ensures
that we sum over updated $\delta$-GLMB/LMB global hypotheses that
assign measurement $z$ to one of the targets, $w_{k}^{\left(I_{+},\theta\right)}$
is the weight of the updated global hypothesis that contains a set
of labels $I_{+}$ and data associations $\theta$. Details can be
found in \cite{Reuter14}. 

Filtering labelled RFS with adaptive birth model has these properties:

\begin{enumerate}
\item It can improve the estimated trajectories by $\delta$-GLMB/LMB filters
for an IID cluster birth process covering a large area, see Figure
\ref{fig:Track_switching}.  
\item Target estimation at birth time is delayed at least one time step,
see Figure \ref{fig:Track_switching}.
\item The filter no longer has information on undetected targets, which
is required, for example, in sensor management \cite{Bostrom-Rost21}.
\item Target birth depends on past measurements and the $\delta$-GLMB/LMB
filter output.
\item The $\delta$-GLMB filter is no longer a closed-form recursion to
obtain the posterior, as Property 4 does not meet the $\delta$-GLMB
filter derivation assumptions, which require that target birth is
independent of other targets and measurements.
\item Contrary to standard Bayesian filtering modelling \cite{Sarkka_book13},
there is no generative model of the random variables that represent
the states and the measurements that is independent of the filtering
algorithm. Therefore, we cannot simulate the ground truth set of targets
and measurements by drawing samples from the random process, and,
afterwards, carry out filtering.
\end{enumerate}

\section{Bernoulli track initiation from isolated measurements}

In this section, we compare adaptive birth and PMBM Bernoulli track
initiation by an isolated measurement that lies in a region where
there are no previously detected targets present. We consider this
type of scenario as it is easy to analyse and gives insights in the
differences. We observe $Z_{k}=\left\{ z_{k}^{1},...,z_{k}^{m_{k}}\right\} $
and consider the assumptions
\begin{itemize}
\item A1 Measurement $z_{k}^{1}$ is far away from all potential targets
that have been previously detected according to the filter.
\item A2 Measurements $z_{k}^{2},...,z_{k}^{m_{k}^{u}}$, $m_{k}^{u}\leq m_{k}$
are far away from $z_{k}^{1}$ and all potential targets that have
been previously detected according to the filter.
\end{itemize}
Sections \ref{subsec:PMBM-Bernoulli-initialisation} and \ref{subsec:Adaptive-birth-Bernoulli}
explain Bernoulli initiation for PMBM and adaptive birth, respectively.
Section \ref{subsec:Illustration-of-differences} illustrates the
differences between both methods in three cases.

\subsection{PMBM Bernoulli initiation\label{subsec:PMBM-Bernoulli-initialisation}}

Due to A1, the new Bernoulli created by $z_{k}^{1}$ appears in all
updated global hypotheses of the PMBM that have non-zero weight. The
updated probability of existence and single-target density are given
by (\ref{eq:r_k_PMBM}) and (\ref{eq:p_k_PMBM}). 

As adaptive birth creates the new Bernoulli components at the following
time step, we must perform a prediction on the Bayesian Bernoulli
to compare both approaches at the same time step. After the prediction,
the probability of existence and single-target density of the Bernoulli
are \cite{Williams15b}
\begin{align}
r_{k+1|k} & =\left\langle p_{k|k},p^{S}\right\rangle \label{eq:r_k_pred_PMBM}\\
p_{k+1|k}\left(x\right) & =\frac{\int g\left(x|y\right)p^{S}\left(y\right)p_{k|k}\left(y\right)dy}{r_{k+1|k}}.\label{eq:p_k_pred_PMBM}
\end{align}

The Bayesian approach takes into account $p^{D}\left(\cdot\right),\lambda^{C}\left(\cdot\right)$,
$l\left(z_{k}^{1}|\cdot\right)$ and $\lambda_{k|k-1}\left(\cdot\right)$
to obtain (\ref{eq:r_k_PMBM}) and (\ref{eq:p_k_PMBM}), and $p^{S}\left(\cdot\right)$
and $g\left(\cdot|\cdot\right)$ to obtain (\ref{eq:r_k_pred_PMBM})
and (\ref{eq:p_k_pred_PMBM}). In addition, the parameters of the
Bernoulli are independent of other measurements or previous potential
targets.

\subsection{Adaptive birth Bernoulli initiation\label{subsec:Adaptive-birth-Bernoulli}}

Due to A1 and A2, in adaptive birth, the Bernoulli birth components
created by measurements $z\in\left\{ z_{k}^{1},...,z_{k}^{m_{k}^{u}}\right\} $
have $r_{U,k}\left(z\right)=0$. The probability of existence of the
Bernoulli created by $z_{k}^{1}$ is then
\begin{align}
 & \hat{r}_{B,k+1}^{(\ell)}\left(z_{k}^{1}\right)\nonumber \\
 & =\min\left(r_{B,max},\frac{\overline{\lambda}_{k+1}^{B}}{m_{k}^{u}+\sum_{j=m_{k}^{u}+1}^{m_{k}}\left(1-r_{U,k}\left(z_{k}^{j}\right)\right)}\right),\label{eq:adaptive_far_away_existence}
\end{align}
and its single target density $\hat{p}_{k+1|k}\left(x\right)$ is
user-defined. 

Contrary to the Bayesian Bernoulli initiation, the adaptive birth
probability of existence (\ref{eq:adaptive_far_away_existence}) depends
on the events in other areas through $m_{k}^{u}$ and the sum in the
denominator. That is, the probability of existence of a Bernoulli
created by an isolated measurement is affected by all measurements
in the scene. This happens even if they are far-away and do not contain
information on this potential target. This can be considered as a
type of spooky action at a distance \cite{Franken09}. 

It should also be noted that, in contrast to the Bayesian approach,
adaptive birth does not weight the hypotheses that this measurement
has been generated by clutter or by a new detection. In addition,
if $m_{k}=m_{k}^{u}$, the probability of existence does not depend
on the parameters used in the Bayesian formulation: $p^{D}\left(\cdot\right)$,
$\lambda^{C}\left(\cdot\right)$, $l\left(z_{k}^{1}|\cdot\right)$,
$\lambda_{k|k-1}\left(\cdot\right)$, $p^{S}\left(\cdot\right)$ and
$g\left(\cdot|\cdot\right)$.

\subsection{Illustration of differences in three scenarios\label{subsec:Illustration-of-differences}}

We illustrate how both methods behave in three cases in which differences
arise. All cases consider $\lambda_{k|k-1}\left(x\right)>0$ and $p^{D}\left(x\right)>0$.

\subsubsection{Case 1}

We consider a scenario without clutter in the area of $z_{k}^{1}$.
For $\lambda^{C}\left(z_{k}^{1}\right)=0$, $p^{S}\left(x\right)=p^{S}$,
the Bayesian approach sets
\begin{equation}
r_{k|k}=1,\;r_{k+1|k}=p^{S},\label{eq:r_k_Bayesian_1}
\end{equation}
as this measurement cannot have been originated by clutter. Instead,
adaptive birth can set $\hat{r}_{B,k+1}^{(\ell)}$ arbitrarily low
if $m_{k}^{u}\rightarrow\infty$, which may happen if the clutter
intensity or number of targets is high in other areas without previously
detected targets. 

\subsubsection{Case 2}

We consider a scenario in which there is a large number of targets
in a far away area with high probability of detection, and $\lambda^{C}\rightarrow0$.
In this case, we expect a large number $m_{k}^{u}$ of received measurements,
originated from other targets. The Bayesian solution also sets (\ref{eq:r_k_Bayesian_1}),
as it is known that $z_{k}^{1}$ was generated by a new target. In
contrast, adaptive birth provides $\hat{r}_{B,k+1}^{(\ell)}\left(z_{k}^{1}\right)\rightarrow0$
for $m_{k}^{u}\rightarrow\infty$. 

\subsubsection{Case 3\label{sec:Example-pd_x}}

We analyse another example with a state-dependent probability of detection
and non-uniform clutter. At time step 0, there are no targets with
probability one, and we analyse the predicted information at time
step 2.

We consider static targets, which could represent landmarks in robotics
or mapping \cite{Fatemi17,Durrant06}. The target state is $x=\left[x_{1},x_{2}\right]^{T}$
where $x_{1}$ is position in the $x$-axis and $x_{2}$ is position
in the $y$-axis. The target birth intensities $\lambda_{1}^{B}\left(\cdot\right)$
and $\lambda_{2}^{B}\left(\cdot\right)$ are uniform in the surveillance
area $\left[-l_{x},l_{x}\right]\times\left[-l_{x},l_{x}\right]$,
with $l_{x}=10\,\mathrm{m}$. 

\begin{figure}
\begin{centering}
\subfloat[]%
{\begin{centering}
\includegraphics[scale=0.5]{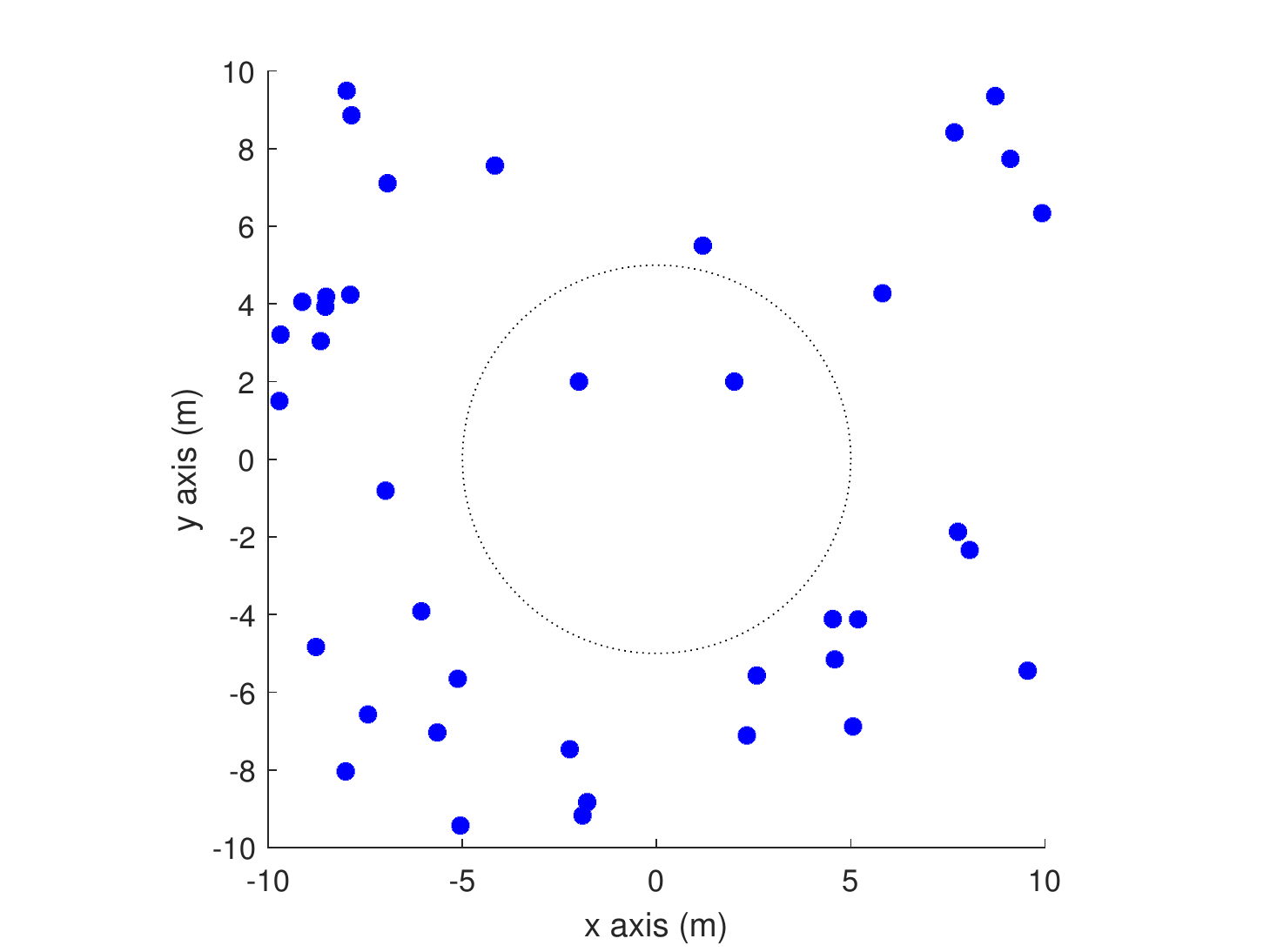}
\par\end{centering}
}
\par\end{centering}
\begin{centering}
\subfloat[]%
{\begin{centering}
\includegraphics[scale=0.5]{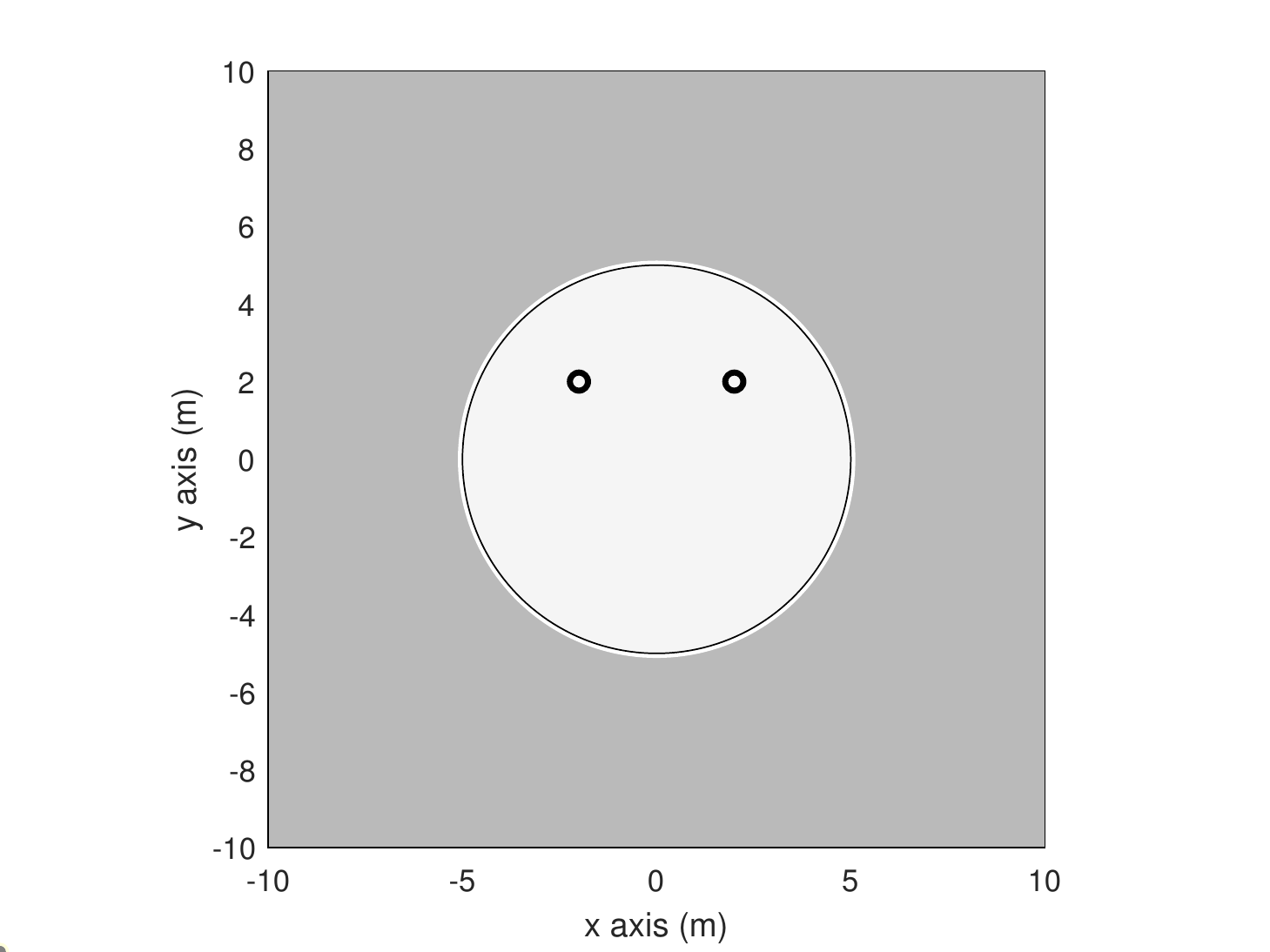}
\par\end{centering}
}
\par\end{centering}
\begin{centering}
\subfloat[]%
{\begin{centering}
\includegraphics[scale=0.5]{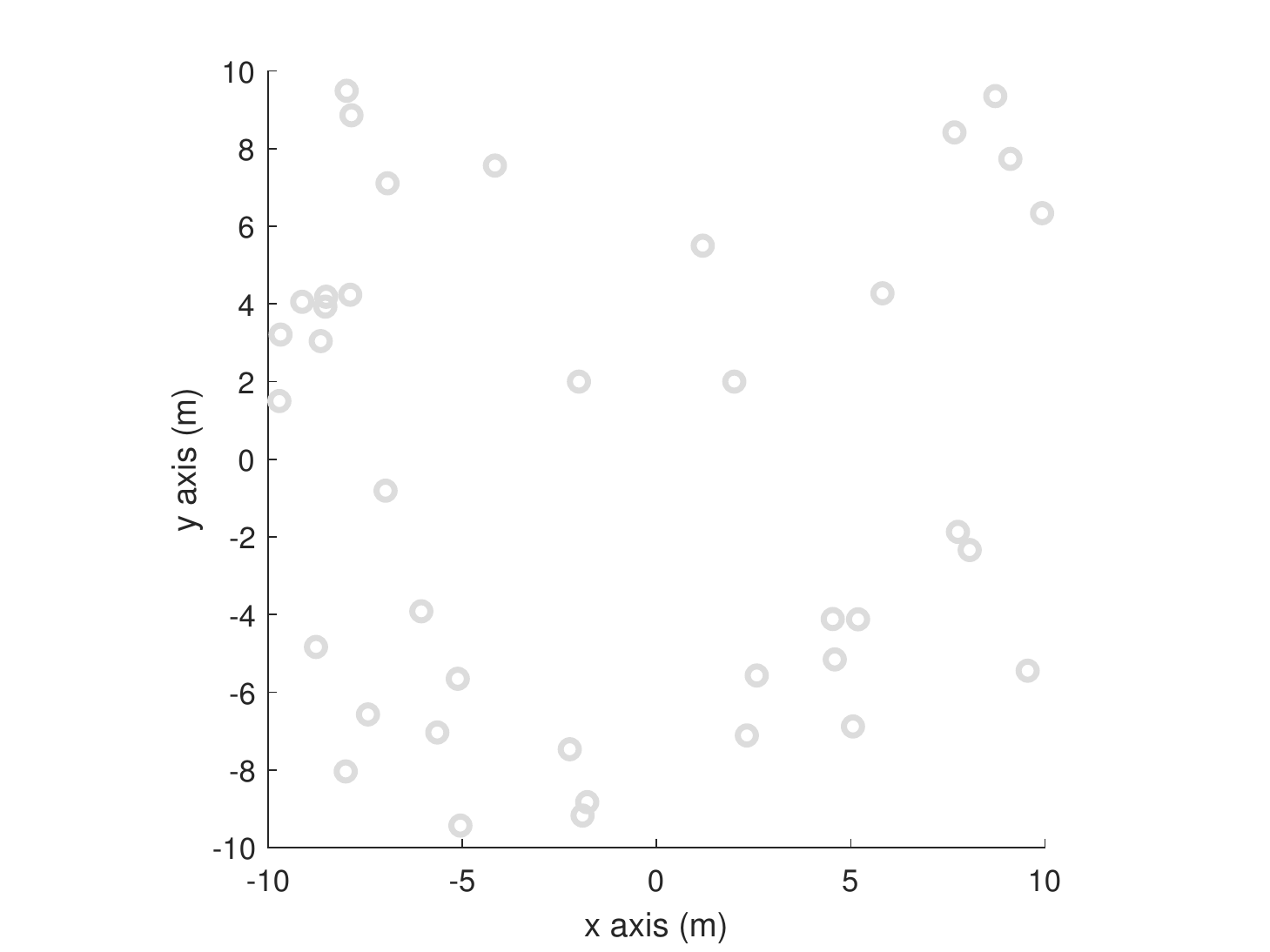}
\par\end{centering}
}
\par\end{centering}
\caption{\label{fig:State-dependent_pd}Illustration of Bayesian versus adaptive
birth predicted densities $p^{S}=1$: $p^{D}(x)=1$ and $\lambda^{C}(z)=0$
inside the circle, $p^{D}(x)=0$ and $\lambda^{C}(z)>0$ outside the
circle. There are two target generated measurements and 37 clutter
measurements. Subfigure (a): measurements at time step 1. Subfigure
(b): in the Bayesian solution, there may be unobserved targets outside
the FoV, new born targets inside and outside the FoV, and there are
also two targets (black circles) inside the FoV with probability one.
Subfigure (c): the adaptive birth solution indicates there are 39
potential targets (grey circles), each with a probability of existence
$\min\left(r_{B,max},\frac{\overline{\lambda}_{2}^{B}}{39}\right)$,
where $r_{B,max}$ is user-defined. In Subfigures (b) and (c), darker
areas represent a higher probability of potential targets.}
\end{figure}

The probability of detection is 1 in the field of view (FoV), which
is a circle of radius $R^{D}=5\,\mathrm{m}$, such that
\begin{align}
p^{D}(x) & =\begin{cases}
1 & \left\Vert x\right\Vert \leq R^{D}\\
0 & \left\Vert x\right\Vert >R^{D}.
\end{cases}
\end{align}
The single-measurement density for a target-generated measurement
$z=\left[z_{1},z_{2}\right]^{T}$ is $l\left(z|x\right)=\mathcal{N}\left(z;x,\alpha I\right)$,
with $\alpha>0$ being a small number. The clutter intensity is

\begin{align}
\lambda^{C}(z) & =\begin{cases}
0 & \left\Vert z\right\Vert \leq R^{D}\\
\lambda^{C} & R^{D}<\left\Vert z\right\Vert ,\left\Vert z_{1}\right\Vert <l_{x},\left\Vert z_{2}\right\Vert <l_{x}
\end{cases}
\end{align}
where $\lambda^{C}>0$. Therefore, in this scenario, we have almost
perfect detections with no clutter in the circle located at the origin
with radius $R^{D}$. Outside the circle, we cannot detect potential
targets and we only have clutter measurements. 

Let us consider we observe the measurements in Figure \ref{fig:State-dependent_pd}(a).
The available information in a Bayesian filter, shown in Figure \ref{fig:State-dependent_pd}(b),
is that, at time step 2, the two detections inside the circle are
target detections with probability one. Then, there may be undetected
targets outside the circle, and we also have target births at time
step 2 in the whole surveillance area. On the contrary, the adaptive
birth solution in Figure \ref{fig:State-dependent_pd}(c) does not
make use of the knowledge on how $p^{D}$ and $\lambda^{C}$ change
spatially, though this information is actually used in the filters
to perform prediction and update. The filter incorrectly thinks that
there is a potential target with existence probability $\min\left(r_{B,max},\frac{\overline{\lambda}_{2}^{B}}{39}\right)$
created by each of the 39 measurements. 

\section{Expected cardinality analysis\label{sec:Analysis-predicted_cardinality}}

This section analyses the expected number of targets using Bayesian
and adaptive birth. To obtain closed-form equations, we analyse the
predicted cardinality at time step 2 averaged over all measurements
under these conditions
\begin{itemize}
\item C1 At time step 0, there are no targets in the surveillance area with
probability 1.
\item C2 $p^{S}$ and $p^{D}$ are constants that do not depend on the target
state.
\end{itemize}
The analysis of the cardinality at time step 1 is not of high interest,
as adaptive birth delays target births by one time step, so we consider
time step 2. We first provide the Bayesian solution in Section \ref{subsec:Bayesian-solution_average_cardinality}.
Then, we analyse the adaptive birth solution in Section \ref{subsec:Adaptive-birth-solution_average_cardinality}.

\subsection{Bayesian solution\label{subsec:Bayesian-solution_average_cardinality}}

The MTT system starts at time step 0 with C1, then, targets may be
born at time step 1 generating measurements at time step 1 according
to the models in Section \ref{subsec:Bayesian-filtering-recursions}.
At time step 2, new targets may be born. If we average over all possible
measurements at time step 1, the predicted number of targets at time
step 2 is
\begin{align}
E\left[|X_{2}|\right] & =\int\int|X_{2}|p(X_{2},Z_{1})\delta X_{2}\delta Z_{1}\\
 & =\int|X_{2}|p(X_{2})\delta X_{2}\\
 & =p^{S}\overline{\lambda}_{1}^{B}+\overline{\lambda}_{2}^{B}\label{eq:predicted_cardinality_bayesian}
\end{align}
where $p(X_{2},Z_{1})$ is the joint density of $X_{2}$ and $Z_{1}$,
and $|X|$ denotes the cardinality of set $X$. That is, the expected
number of targets at time step 2 is the expected number of targets
born at time step 1 that have survived plus the expected number of
targets born at time step 2. Equation (\ref{eq:predicted_cardinality_bayesian})
can also be obtained by integrating the predicted PHD in the PHD filter
\cite{Mahler_book14}.

\subsection{Adaptive birth solution\label{subsec:Adaptive-birth-solution_average_cardinality}}

As there are no targets at time step 0, $r_{U,1}\left(z\right)=0$
in (\ref{eq:r_u}) for any measurement $z\in Z_{1}$. This implies
that, for $z\in Z_{1}$, (\ref{eq:r_b}) becomes
\begin{align}
\hat{r}_{B,2}^{(\ell)}\left(z\right) & =\min\left(r_{B,max},\frac{\overline{\lambda}_{2}^{B}}{|Z_{1}|}\right).\label{eq:r_b_time1}
\end{align}
Therefore, for adaptive birth, the predicted number of targets at
time step 2 when we receive |$Z_{1}|$ measurements is
\begin{align}
\hat{N}(|Z_{1}|) & =\min\left(|Z_{1}|r_{B,max},\overline{\lambda}_{2}^{B}\right).\label{eq:expected_number_targets_adaptive}
\end{align}

Then, the expected number of targets averaged over all measurements
at time step 2 is
\begin{align}
\hat{E}\left[|X_{2}|\right] & =\int\int|X_{2}|p(X_{2},Z_{1})\delta X_{2}\delta Z_{1}\nonumber \\
 & =\int\hat{N}(|Z_{1}|)p(Z_{1})\delta Z_{1}\\
 & =\sum_{m=0}^{\infty}\min\left(mr_{B,max},\overline{\lambda}_{2}^{B}\right)\rho_{Z_{1}}\left(m\right)\\
 & =r_{B,max}\sum_{m=0}^{\check{m}}m\rho_{Z_{1}}\left(m\right)+\overline{\lambda}_{2}^{B}\sum_{m=\check{m}+1}^{\infty}\rho_{Z_{1}}\left(m\right)\\
 & =\overline{\lambda}_{2}^{B}+\sum_{m=0}^{\check{m}}\left(r_{B,max}m-\overline{\lambda}_{2}^{B}\right)\rho_{Z_{1}}\left(m\right)
\end{align}
where $\rho_{Z_{1}}\left(m\right)$ is the probability that set $Z_{1}$
contains $m$ elements and
\begin{align*}
\check{m} & =\left\lfloor \overline{\lambda}_{2}^{B}/r_{B,max}\right\rfloor .
\end{align*}
With adaptive birth model, the distribution $\rho_{Z_{1}}\left(\cdot\right)$
is not specified as there is no generative model for the random variables. 

We should note that the expected number of targets of the adaptive
birth model meets
\begin{align}
\sum_{m=0}^{\infty}\min\left(mr_{B,max},\overline{\lambda}_{2}^{B}\right)\rho_{Z_{1}}\left(m\right) & \leq\sum_{m=0}^{\infty}\overline{\lambda}_{2}^{B}\rho_{Z_{1}}\left(m\right)\label{eq:Cardinality_bias_bound1}\\
 & =\overline{\lambda}_{2}^{B}\nonumber 
\end{align}
Therefore, the expected number of targets of adaptive birth at time
step 2 is always lower or equal than the Bayesian expected number
of targets.  This result is summarised in the following lemma.
\begin{lem}
\label{lem:Cardinality}Under C1 and C2, the predicted number $\hat{E}\left[|X_{2}|\right]$
of targets at time step 2 of adaptive birth is lower or equal than
the Bayesian solution $E\left[|X_{2}|\right]$
\begin{align}
\hat{E}\left[|X_{2}|\right] & \leq E\left[|X_{2}|\right].
\end{align}
The inequality is strict if $p^{S}\overline{\lambda}_{1}^{B}>0$ or
there is $m\in\mathbb{N}$ such that $mr_{B,max}<\overline{\lambda}_{2}^{B}$
and $\rho_{Z_{1}}\left(m\right)>0$. In addition, the value of $r_{B,max}\in\left[0,1\right]$
that minimises the cardinality bias gap $E\left[|X_{2}|\right]-\hat{E}\left[|X_{2}|\right]$
is $r_{B,max}^{*}=1$.
\end{lem}
As the estimated number of targets with adaptive birth is lower than
with Bayesian birth under mild conditions, see Lemma \ref{lem:Cardinality},
we will generally expect that filters with adaptive births miss more
targets than filters with Bayesian birth.

\section{Simulations}

We compare the performance of Bayesian and adaptive birth via simulations.
We have implemented the $\delta$-GLMB and LMB filters\footnote{Matlab code is available at http://ba-tuong.vo-au.com.}
with joint prediction and update, using Murty's algorithm \cite{Murty68},
with and without adaptive birth. The versions with adaptive birth
are referred to as A-$\delta$-GLMB and A-LMB. These filters have
been implemented with a maximum number of global hypotheses equal
to 1000, and pruning threshold $10^{-10}$. LMB has been implemented
propagating a single Gaussian, merging threshold 4 and Bernoulli pruning
threshold $10^{-3}$. 

We have implemented the PMBM and PMB filters \cite{Williams15b} with
Murty's algorithm\footnote{Matlab code is available at https://github.com/Agarciafernandez/MTT.}.
We have also implemented adaptive birth, following Section \ref{subsec:Adaptive-birth-model},
with the multi-Bernoulli mixture (MBM) filter \cite{Angel18_b,Angel19_e}
and multi-Bernoulli (MB) filter. The adaptive MBM (A-MBM) and adaptive
MB (A-MB) filters correspond to the PMBM and PMB filters but with
the adaptive multi-Bernoulli birth (setting the Poisson intensity
equal to zero). This implies that the main difference between A-MBM
and PMBM, and A-MB and PMB is the way tracks (Bernoulli components)
are initiated. These filters have been implemented with the following
parameters \cite{Angel18_b}: maximum number of global hypotheses
$N_{h}=200$, threshold for pruning the PPP weights $\Gamma_{p}=10^{-5}$,
threshold for pruning Bernoulli components $\Gamma_{b}=10^{-5}$,
estimator 1 with threshold 0.4, and ellipsoidal gating with threshold
20. To speed up running times, all the filters have been implemented
with the compiled Murty's algorithm in \cite{Crouse17}. All units
in this section are in the international system. 

\subsection{Models}

A target state consists of position and velocity $[p_{x},v_{x},p_{y},v_{y}]^{T}$
with a nearly constant velocity model 
\begin{align*}
g\left(x_{k}|x_{k-1}\right) & =\mathcal{N}\left(x_{k};Fx_{k-1},Q\right)
\end{align*}
\[
F=I_{2}\otimes\begin{bmatrix}1 & T\\
0 & 1
\end{bmatrix},\,Q=qI_{2}\otimes\begin{bmatrix}T^{3}/3 & T^{2}/2\\
T^{2}/2 & T
\end{bmatrix},
\]
where $\otimes$ is the Kronecker product, $q=0.01$, and the sampling
time $T=1$. We also consider $p_{S}=0.995$. 

The sensor measures target positions with the model
\begin{align*}
l\left(z|x\right) & =\mathcal{N}\left(z;Hx,R\right)
\end{align*}
\[
H=I_{2}\otimes\begin{bmatrix}1 & 0\end{bmatrix},\,R=I_{2}.
\]
Clutter is uniformly distributed in the region of interest $A=[0,1000]\times[0,1000]$
with intensity $\lambda^{C}\left(z\right)=\overline{\lambda}^{C}\cdot u_{A}\left(z\right)$,
where $u_{A}\left(z\right)$ is a uniform density and $\overline{\lambda}^{C}=10$.
The probability of detection is $p_{D}=0.9$.

All filters assume that there are no targets at time 0. Filters with
PPP birth have a Gaussian intensity for new born targets with mean
$\overline{x}_{k}^{b,1}=\left[100,0,100,0\right]^{T}$ and covariance
matrix $P_{k}^{b,1}=\mathrm{diag}\left(\left[150^{2},1,150^{2},1\right]\right)$,
with weight $w_{1}^{b,1}=10$ and $w_{k}^{b,1}=0.1$ for $k>1$. The
$\delta$-GLMB and LMB filters use 18 Bernoulli birth components,
each with probability of existence $w_{1}^{b,1}/18$ at time step
1. From time step 2, the birth has one Bernoulli component with probability
of existence 0.1. The spatial density of these Bernoulli birth components
is the same as for the PPP birth. We use 18 Bernoulli components at
time step 1 so that the multi-Bernoulli birth is a reasonable approximation
of the PPP birth. The probability of existence is set so that the
PPP birth and multi-Bernoulli birth have the same intensity \cite{Mahler_book14}.
For adaptive birth, we use (\ref{eq:r_b}) with $\overline{\lambda}_{k}^{B}=r_{B,max}=w_{k}^{b,1}$.
The user-defined single target-density for adaptive birth is the one
in \cite{Reuter14}. The mean state is the position indicated by the
measurement with zero velocity, and the covariance matrix is $100I_{4}.$

\begin{figure}
\begin{centering}
\includegraphics[scale=0.6]{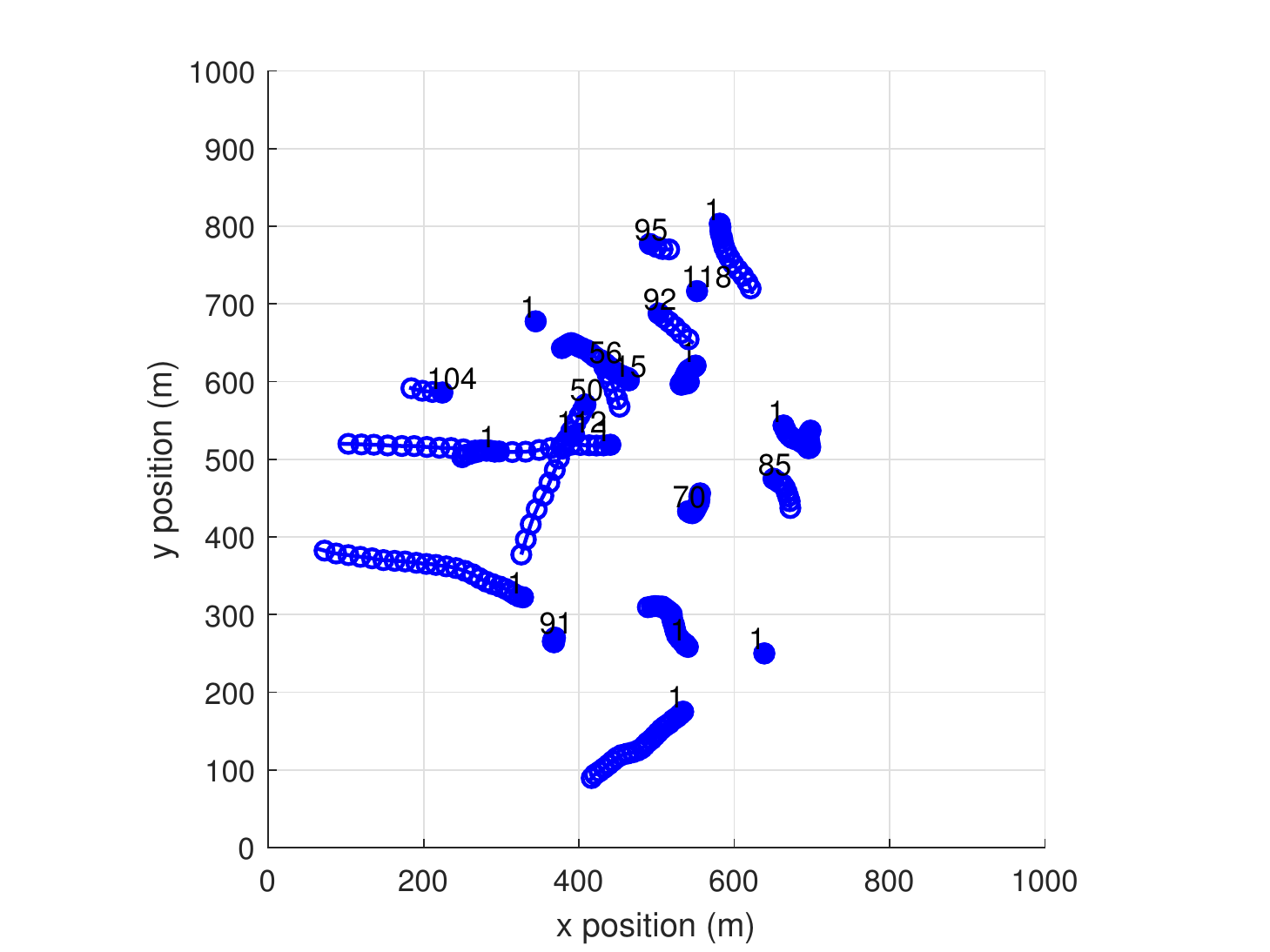}
\par\end{centering}
\begin{centering}
\includegraphics[scale=0.6]{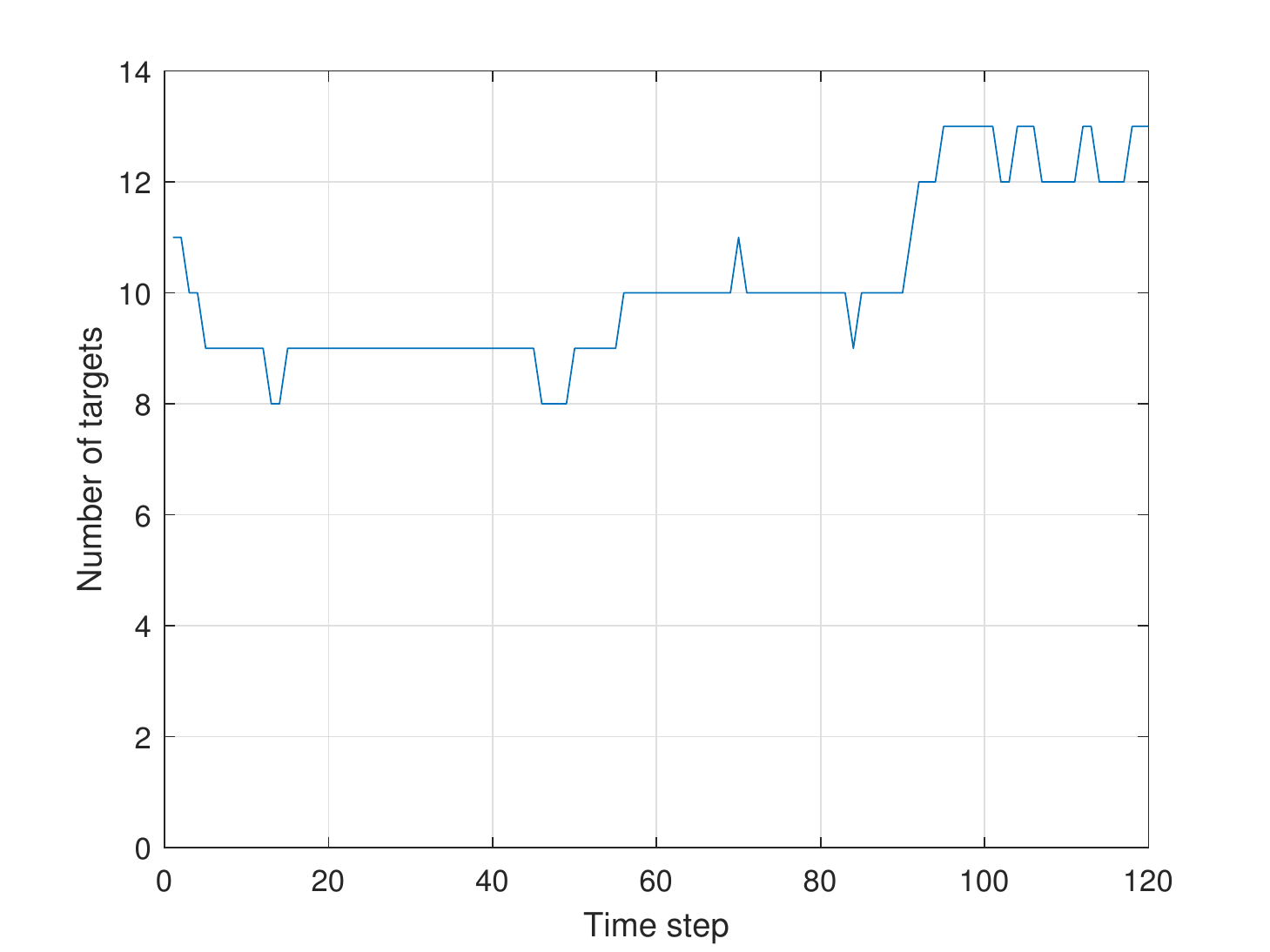}
\par\end{centering}
\caption{\label{fig:Scenario-simulations}Scenario of the simulations: set
of trajectories (top) and number of targets present at each time step
(bottom). The number next to each trajectory is its time of birth.
The total number of trajectories is 22.}
\end{figure}

We draw the ground truth set of trajectories from the generative model
with PPP birth and 120 time steps. The resulting sets of trajectories
are shown in Figure \ref{fig:Scenario-simulations}. 

\subsection{Simulation results}

\begin{figure}
\begin{centering}
\includegraphics[scale=0.3]{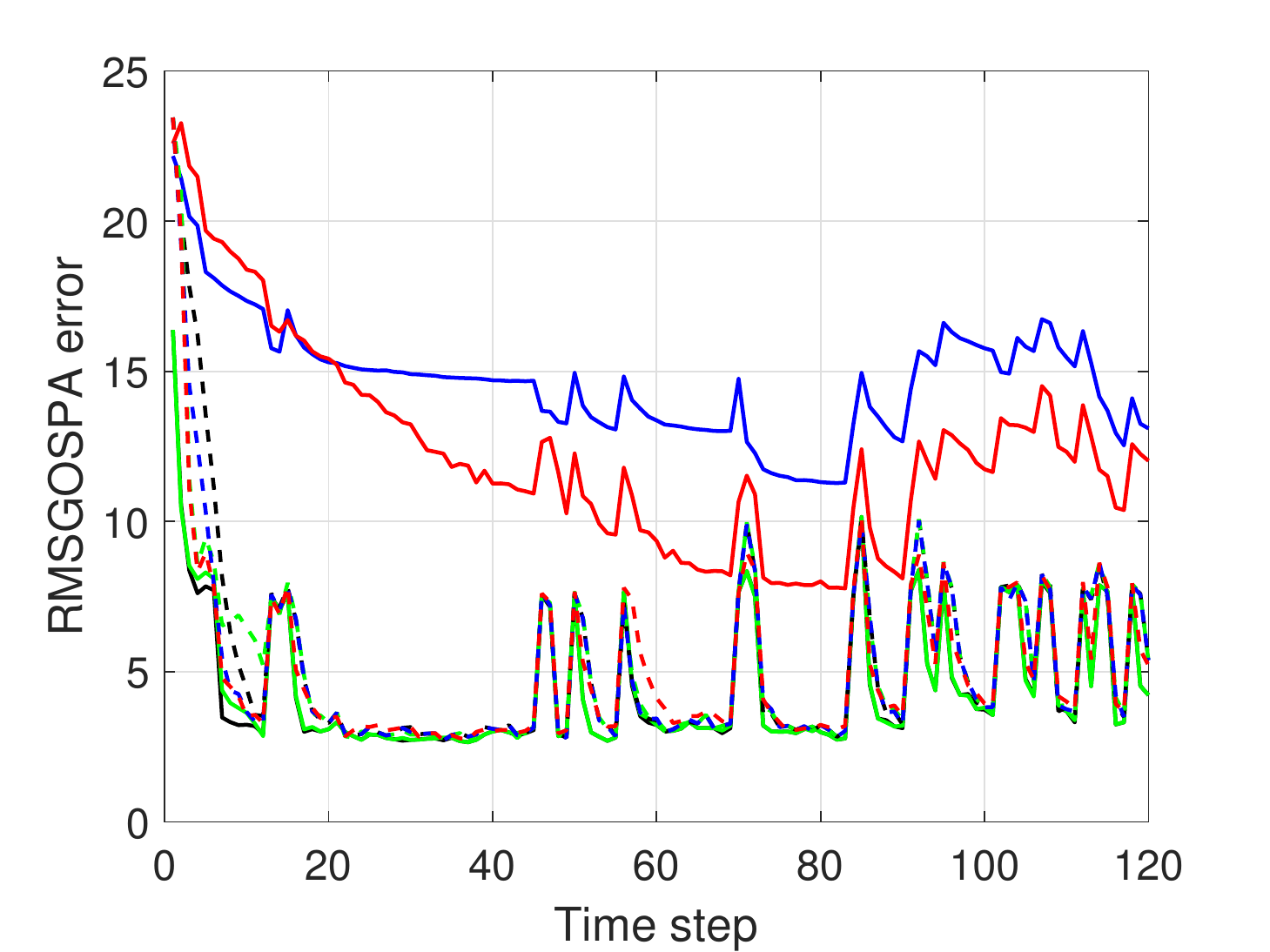}\includegraphics[scale=0.3]{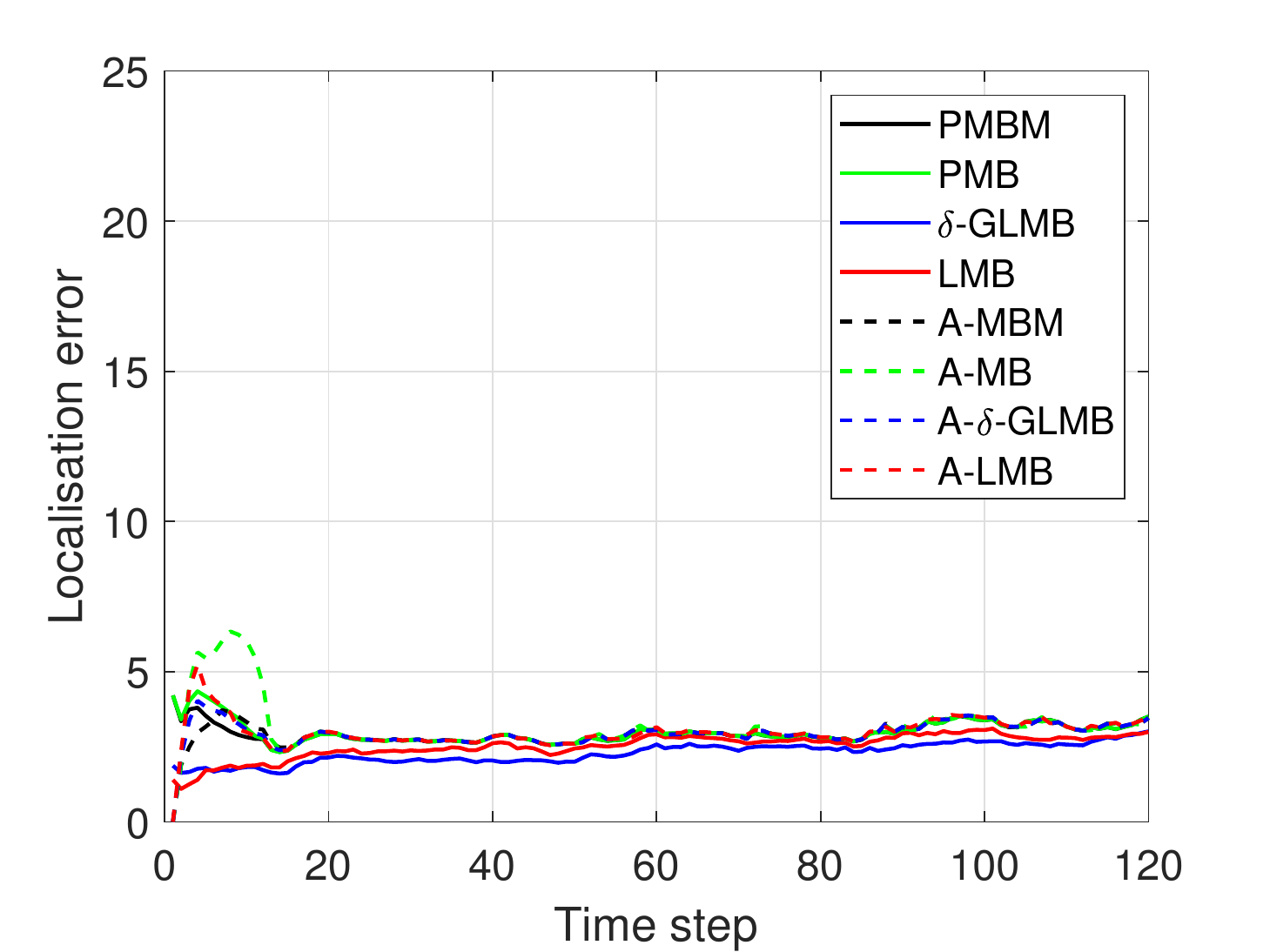}
\par\end{centering}
\begin{centering}
\includegraphics[scale=0.3]{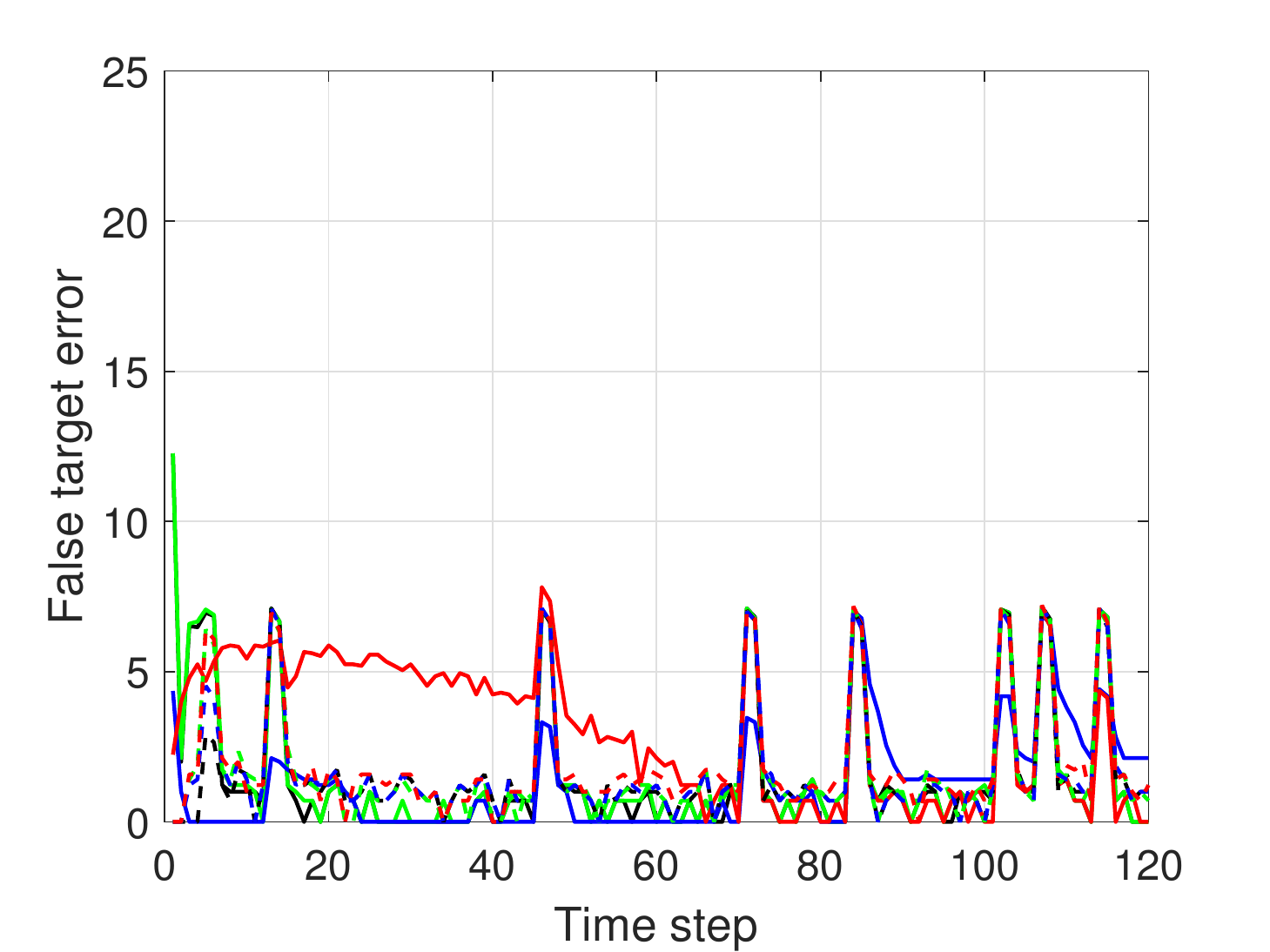}\includegraphics[scale=0.3]{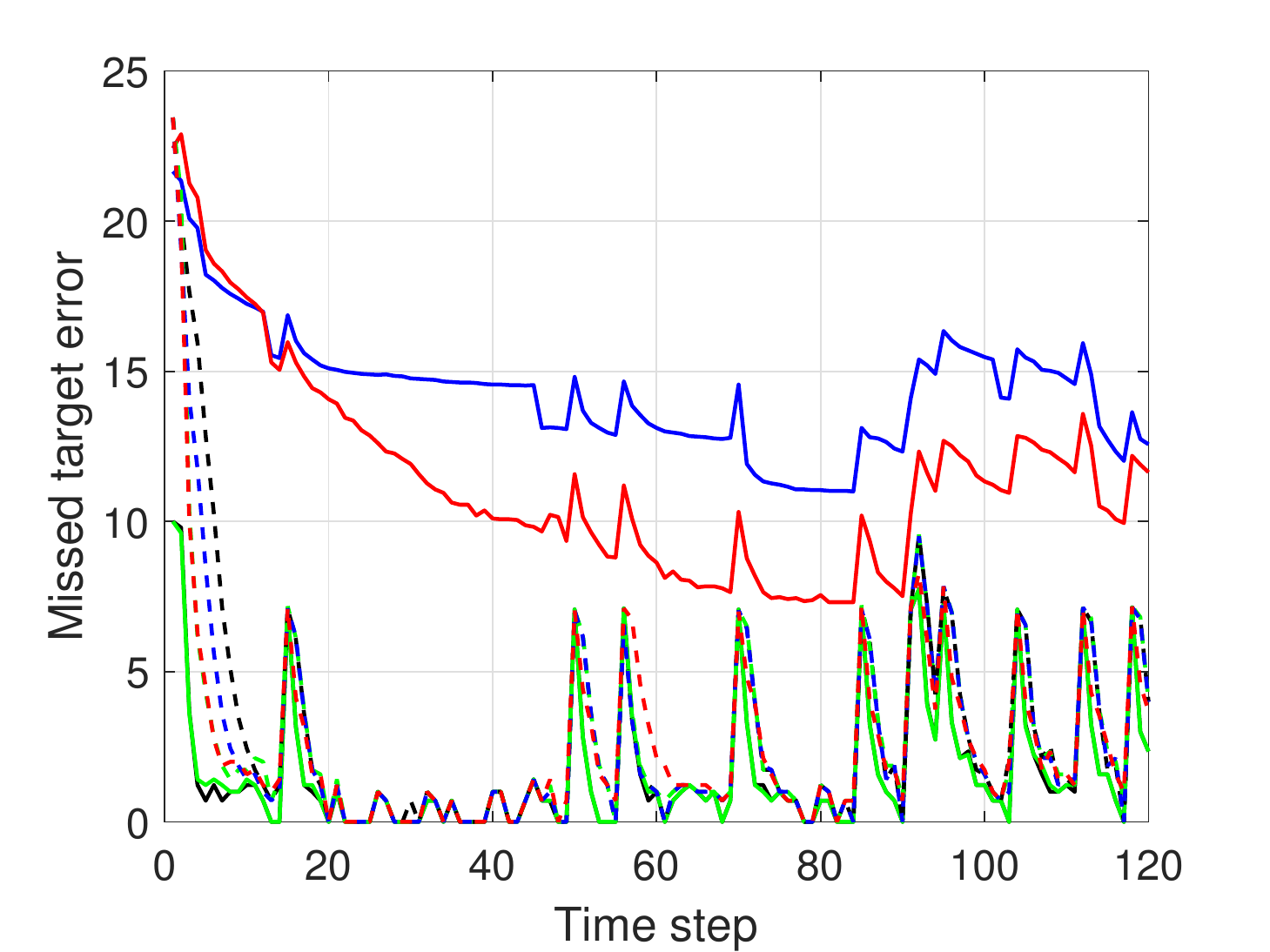}
\par\end{centering}
\caption{\label{fig:RMS-GOSPA-errors}RMS GOSPA errors and their decomposition
against time for the considered filters. }

\end{figure}

We assess filter performance via Monte Carlo simulation with 100 runs
and obtain the root mean square generalised optimal subpattern assignment
(RMS-GOSPA) metric error ($p=2$, $c=10$, $\alpha=2$) \cite{Rahmathullah17}.
The resulting errors as well as the metric decomposition into localisation
error for properly detected targets, and costs for missed and false
targets are shown in Figure \ref{fig:RMS-GOSPA-errors}. The best
performing filters are the PMBM and PMB filters. These are followed
by the adaptive birth filters: A-$\delta$-GLMB, A-LMB, A-MBM and
A-MB. The $\delta$-GLMB and LMB filters achieve considerably worse
performance. The reason for this performance is the high number of
hypotheses that are required by $\delta$-GLMB and LMB to keep relevant
information, specially when target birth is an IID cluster process
covering a large area and more than one target may be born at the
same time step \cite[App. D]{Angel20_e}. Adaptive birth filters tend
to miss more targets than the PMBM and PMB filters when targets are
born. 

The computational times to execute one Monte Carlo run of the algorithms
on an Intel core i5 laptop are given in Table \ref{tab:Computational-time}.
The PMB filter is the fastest algorithm followed by A-MB. These are
followed by the PMBM and A-MBM filters. The $\delta$-GLMB and LMB
filters are slower due to the MBM$_{01}$ expansion required in each
prediction-update stage and because we consider a higher number of
global hypotheses. 

The LMB filter is slower than the $\delta$-GLMB filter in this scenario.
In the profile, we can see that the joint prediction and update is
faster in LMB, but the function that projects the updated $\delta$-GLMB
into an LMB takes considerable time. The $\delta$-GLMB and LMB filters
without adaptive birth are faster than with adaptive birth as they
consider a maximum of one new born target from time step 2, and therefore
fewer hypotheses.

On the whole, we can conclude that filters that do not require an
MBM$_{01}$ expansion have important computational benefits. In addition,
filters with Bayesian birth are more accurate. Overall PMBM and PMB
are the best performing filters in accuracy and computational time.

\begin{table*}
\caption{\label{tab:Computational-time}Computational times in seconds ($\overline{\lambda}^{C}=10$)}

\begin{centering}
\begin{tabular}{c|cccccccc}
 &
PMBM &
PMB &
$\delta$-GLMB &
LMB &
A-MBM &
A-MB &
A-$\delta$-GLMB &
A-LMB\tabularnewline
\hline 
Time &
3.1 &
\uline{1.9} &
12.6 &
23.9 &
9.7 &
2.3 &
49.9 &
46.8\tabularnewline
\hline 
\end{tabular}
\par\end{centering}
\end{table*}

\begin{table*}
\caption{\label{tab:GOSPA}RMS-GOSPA errors across time }

\centering{}%
\begin{tabular}{c|cccccccc}
 &
PMBM &
PMB &
$\delta$-GLMB &
LMB &
A-MBM &
A-MB &
A-$\delta$-GLMB &
A-LMB\tabularnewline
\hline 
$\overline{\lambda}^{C}=10$ &
\uline{5.08} &
5.12 &
14.82 &
12.81 &
6.42 &
6.15 &
6.11 &
5.92\tabularnewline
$\overline{\lambda}^{C}=20$ &
\uline{5.27} &
5.30 &
15.24 &
12.34 &
6.82 &
6.40 &
6.58 &
6.26\tabularnewline
$\overline{\lambda}^{C}=30$ &
\uline{5.51} &
5.57 &
12.83 &
13.95 &
7.04 &
6.51 &
6.86 &
6.43\tabularnewline
\hline 
\end{tabular}
\end{table*}

\section{Conclusion}

In this paper, we have compared the adaptive birth model used in the
labelled RFS framework with the Bernoulli track initiation in PMBM/PMB
filters. Adaptive birth resembles the track initiation in PMBM/PMB
filters, in which each measurement gives rise to a new Bernoulli.
Though not obtained from Bayesian principles, adaptive birth generally
works well in scenarios with constant probability of detection and
clutter intensity. Adaptive birth also improves target and trajectory
estimation for the LMB and $\delta$-GLMB filters when the birth model
is an IID cluster process with large spatial uncertainty and more
than one target may be born at the same time step.  

Simulation results show that PMBM/PMB filters outperform filters with
adaptive birth to estimate the set of targets, both in accuracy and
computational speed. To estimate the set of trajectories, we can use
PMBM/PMB with auxiliary variables that can link information from a
potential target that was first detected by a given measurement \cite{Angel20_e,Granstrom18},
or even better, we can use a PMBM/PMB defined on the set of trajectories
of interest \cite{Angel20_b,Angel20_e}. We therefore argue that it
is preferable to use fully Bayesian MTT filters based on PPP birth
rather than (labelled or unlabelled) multi-Bernoulli adaptive birth. 

\bibliographystyle{IEEEtran}
\bibliography{12C__Trabajo_laptop_Referencias_Referencias}

\end{document}